\documentclass[aps,preprint,preprintnumbers,amsmath,amssymb,UTF8]{revtex4}
\usepackage{amsmath,mathrsfs,amsbsy,color,graphicx,bm,amsthm,amsfonts}
\usepackage{units}
\usepackage{bbm}
\usepackage{times}
\usepackage{dcolumn}
\usepackage{mathrsfs}
\usepackage{amsmath,amssymb,epsfig}
\usepackage{graphicx}
\usepackage{epstopdf}
\usepackage{hyperref}
\usepackage{amssymb}
\usepackage{amsbsy}
\usepackage{graphicx}
\usepackage{amsmath}
\usepackage{times,txfonts}

\newcommand{\gr}[1]{\boldsymbol{#1}}
\newcommand{\sig}{\boldsymbol{\sigma}}
\begin{document}

\title{Black-box estimation of expanding parameter  for de Sitter universe}
\author{ Lulu Xiao$^{1}$, Cuihong Wen$^{1}$\footnote{Email: cuihongwen@hunnu.edu.cn}, Jiliang Jing$^{1}$\footnote{Email: jljng@hunnu.edu.cn}, and Jieci Wang$^{1}$\footnote{Email: jcwang@hunnu.edu.cn}}
\affiliation{$^1$ Department of Physics, and Collaborative Innovation Center for Quantum Effects \\
and Applications,
 Hunan Normal University, Changsha, Hunan 410081, China
}


\begin{abstract}
We study the black-box parameter estimation of expanding parameters and the dynamics of Gaussian interferometric power for the  de Sitter space. We find that the state between  separated open charts can be employed as a probe state for the black-box quantum metrology. This is nontrivial because the open charts are causally disconnected 
and classical information can not be exchanged between them according to the general relativity.  It is shown that the mass of the scalar field remarkably affects the accuracy of the black-box parameter estimation in the de Sitter space, which is quite different from the flat space case where the mass parameter  does not influence the precision of  estimation.   Quantum discord is found to be a key resource for the estimation of the expanding parameter when there is no entanglement between the initially uncorrelated open charts.
It is demonstrated that the role of the probe state between different open charts is quite distinct because  curvature effect of the de sitter space damages quantum resources for the initially correlated probe states, while it generates quantum resources for the initially uncorrelated probe states.

\end{abstract}

\vspace*{0.5cm}

\maketitle
\section{Introduction}
Quantum metrology  \cite{advances} studies the employment of quantum resources (such as entanglement \cite{EN1} and discord \cite{discord1}) on the enhancement of measurement accuracy \cite{Braunstein1994}.
The problem of quantum metrology can be reduced to find optimal scheme for the estimation of parameters that are encoded in a probe system undergoing parameter-dependent
evolutions. Taking into account the laws of quantum mechanics, quantum resources are capable to improve the precision of the gravitational wave detection \cite{Aasi2013,Ma2017,Grote}, distributed quantum sensing~\cite{Guo2020,zhao2021}, quantum imaging \cite{kolobov1999spatial,qi1}, and  clock synchronization~\cite{leibfried2004toward,wangsyn,Rquan}.
In recent years,  quantum metrological technique has been employed to enhance the accuracy of estimation for relativistic effects and spacetime parameters, which gives birth to relativistic quantum metrology \cite{aspachs,jieciRQM1, HoslerKok2013,  RQMuiverse2,detector1,detector2,  RQI6, RQM2, RQM,RQM9,RQM8}. These studies are nontrivial because the estimated parameters govern some key phenomena in which both quantum and spacetime effects are relevant \cite{JHEP1,JHEP2,JHEP3}. Previous researches on the relativistic quantum metrology include the estimation of Unruh-Hawking effect for free modes \cite{aspachs, HoslerKok2013, RQI6}, moving cavities~\cite{RQM,RQM2}, and accelerated detectors \cite{detector1,detector2}, the metrology of   cosmological parameters \cite{jieciRQM1,RQMuiverse2}, and precise measurement of  spacetime parameters of the Earth~\cite{ RQM9, RQM8}, among others.

The advantage of  optical  quantum  metrological settings has been experimentally verified by  the detection of gravitational via squeezed light \cite{Aasi2013,Ma2017,Grote} and the quantum-enhanced clock synchronization \cite{Rquan}. For a customary  optical quantum metrology, the generator of  parameter is known as a priori, then tailored quantum resources such as  squeezed and entangled states can be exploited to improve the precision of
phase estimation. In 2014,  Girolami \emph{et al.} proposed a framework for the black-box quantum parameter estimation \cite{discordGIP}, which can be applied to the case where the generator of the estimated parameter in the probe state is  {\it not known a priori}.
The interferometric power was introduced to quantify the precision for the estimation of encoded parameters in a worst-case scenario \cite{discordGIP}, and has been proved that it is equivalent to a computable measure of discord-type quantum correlations for the probe state. Adesso \cite{GIP} generalized the research on interferometric power to continuous-variable and proposed a closed formula of the Gaussian interferometric power (GIP) for continuous-variable quantum systems. 

In this paper, we suggest a  black-box parameter estimating strategy for the spacetime parameters  and study the dynamics of the GIP for the de Sitter space \cite{Sasaki:1994yt}. The  black-box  estimation  strategy is required because the spacetime effect of the de Sitter space is a two-mode squeezing transformation acting on the causally disconnected regions\cite{Sasaki:1994yt,Kanno16, Albrecht18}. Therefore, it is impossible to perform a  customary metrological process  where the generator of parameter is known as a priori.
In the present work, an optical  Mach-Zender interferometer is involved to estimate the expanding parameter for the de Sitter space.  Such a setup can be theoretically modeled as a dual-arm channel, where the estimated parameter is encoded to one arm only, and the identity operation is applied to the other arm.
 We assume that a subsystem of the initial state is observed by the  global observer Alice who moves along geodesics, the other subsystem is described by the observer Bob who is restricted to one region in the de Sitter space, which is causally disconnected from the other region \cite{Kanno16, Albrecht18}. Under  the spacetime effect of the de Sitter space, the parameters of the  two-mode squeezing transformation are encoded into the black-box device. The information of the spacetime is obtained when the measurements are completed at the output side of the interferometer. We find that the role of the probe state between different open charts is quite distinct for the black-box parameter estimation.   In addition, the quantum discord of probe states serves as  a promising resource for the black-box quantum parameter estimation when there is no entanglement between the initially uncorrelated probe states in the de Sitter space.

The organization of the paper is as follows. In Sec. II we review the solutions of mode functions and Bogoliubov transformations in the de Sitter space. In Sec. III we introduce the scheme of the black-box optical interferometer and the role of the GIP. In Sec. IV we study the black-box estimation of spacetime parameters and the behavior of the GIP in the de Sitter space. In the final section, we summarize our results.
\section{The behavior of  scalar field in the de Sitter space \label{model}}
We consider a free scalar field $\phi$ with mass $m$ which is initially described by two experimenters, Alice and Bob, in the Bunch-Davies vacuum of the de Sitter space. The coordinate frames of the open charts in the de Sitter space can be obtained by analytic continuation from the Euclidean metric. As shown in Fig.1, the spacetime geometry of the de Sitter space is divided into three  open charts, which are denoted by $R$, $L$ and $C$, respectively. We assume that a subsystem of the initial state is observed by the  global observer Alice moving along geodesics, the other subsystem is described by the observer Bob restricted to the region $R$ in the de Sitter space, and the region $R$ is causally disconnected from the region $L$. The metrics for the two causally disconnected open charts $R$ and $L$ in the de Sitter space are given by~\cite{Sasaki:1994yt}
\begin{eqnarray}
ds^2_R&=&H^{-2}\left[-dt^2_R+\sinh^2t_R\left(dr^2_R+\sinh^2r_R\,d\Omega^2\right)
\right]\,,\nonumber\\
ds^2_L&=&H^{-2}\left[-dt^2_L+\sinh^2t_L\left(dr^2_L+\sinh^2r_L\,d\Omega^2\right)
\right]\,,
\end{eqnarray}
where $d\Omega^2$ is the metric on the two-sphere and $H^{-1}$ is the Hubble radius.

The solutions of the Klein-Gordon equation in different regions are found to be
\begin{eqnarray}\label{solutions1}
u_{\sigma p\ell m}(t_{R(L)},r_{R(L)},\Omega)&\sim&\frac{H}{\sinh t_{R(L)}}\,
\chi_{p,\sigma}(t_{R(L)})\,Y_{p\ell m} (r_{R(L)},\Omega)\,,\qquad \nonumber\\
-{\rm\bf L^2}Y_{p\ell m}&=&\left(1+p^2\right)Y_{p\ell m}\,,
\end{eqnarray}
where  $Y_{p\ell m}$ are harmonic functions on the three-dimensional hyperbolic space. In Eq. (\ref{solutions1}), $\chi_{p,\sigma}(t_{R(L)})$ are positive frequency mode functions supporting  on the $R$ and $L$ regions ~\cite{Sasaki:1994yt}
\begin{eqnarray}
\chi_{p,\sigma}(t_{R(L)})=\left\{
\begin{array}{l}
\frac{e^{\pi p}-i\sigma e^{-i\pi\nu}}{\Gamma(\nu+ip+\frac{1}{2})}P_{\nu-\frac{1}{2}}^{ip}(\cosh t_R)
-\frac{e^{-\pi p}-i\sigma e^{-i\pi\nu}}{\Gamma(\nu-ip+\frac{1}{2})}P_{\nu-\frac{1}{2}}^{-ip}(\cosh t_R)
\,,\\
\\
\frac{\sigma e^{\pi p}-i\,e^{-i\pi\nu}}{\Gamma(\nu+ip+\frac{1}{2})}P_{\nu-\frac{1}{2}}^{ip}(\cosh t_L)
-\frac{\sigma e^{-\pi p}-i\,e^{-i\pi\nu}}{\Gamma(\nu-ip+\frac{1}{2})}P_{\nu-\frac{1}{2}}^{-ip}(\cosh t_L)
\,,
\label{solutions}
\end{array}
\right.
\end{eqnarray}
where $\sigma=\pm 1$  distinguish the independent solutions for each open chart
and $P^{\pm ip}_{\nu-\frac{1}{2}}$ are the associated Legendre functions. The  above solutions  can be normalized by the factor
$
N_{p}=\frac{4\sinh\pi p\,\sqrt{\cosh\pi p-\sigma\sin\pi\nu}}{\sqrt{\pi}\,|\Gamma(\nu+ip+\frac{1}{2})|}\,, $ where $p$ is a positive real parameter normalized by $H$. The mass parameter $\nu$ is defined by
$
\nu=\sqrt{\frac{9}{4}-\frac{m^2}{H^2}}\,
$.
Note that the curvature effect starts to appear around $p\sim 1$ in three-dimensional hyperbolic space  \cite{Kanno16, Albrecht18,Sasaki:1994yt,DE1,DE2,DE3}, and the effect of the curvature becomes stronger when $p$ is less than $1$. Therefore, $p$ can be considered as the curvature parameter of the de Sitter space. The mass parameter $\nu$ has two special values:
 $\nu=1/2$ ($m^2=2H^2$) for the conformally coupled massless scalar field, and $\nu=3/2$ for the minimally coupled massless scalar field.

The scalar field can be
 expanded in accordance with the creation and annihilation operators
\begin{eqnarray}
\hat\phi(t,r,\Omega)
=\frac{H}{\sinh t}\int dp \sum_{\ell,m}\phi_{p\ell m}(t)Y_{p\ell m}(r,\Omega)
\,,
\end{eqnarray}
where $a_{\sigma p\ell m}|0\rangle_{\rm BD}=0$ is the  annihilation operator of the Bunch-Davies vacuum, and the Fourier mode field operator $
\phi_{p\ell m}(t)\equiv
\sum_\sigma\left[\,a_{\sigma p\ell m}\,\chi_{p,\sigma}(t)
+a_{\sigma p\ell -m}^\dagger\,\chi^*_{p,\sigma}(t)\right]$ has been
 introduced.
For brevity, the indices $p$, $\ell$, $m$ of $\phi_{p\ell m}$ of the operators and mode functions will be omitted below.

 Since the Fourier mode field operator should
be the same under the change of mode functions, we can relate the operators $(a_\sigma,a_\sigma^\dag)$ and $(b_q,b_q^\dag)$
 by a Bogoliubov transformation\cite{Sasaki:1994yt,Albrecht18,bov3}
\begin{eqnarray}
\phi(t)=a_\sigma\,\chi^\sigma+a_\sigma^\dag\,\chi^\sigma{}^*
=b_q\,\varphi^q+b_q^\dag\,\varphi^q{}^*\,,
\label{fo}
\end{eqnarray}
where  the creation and annihilation operators ($b_q,b_q^\dag$) in different regions are introduced to ensure
 $b_q|0\rangle_{q}=0$. To facilitate the calculation of degrees of freedom in $L$ space, the Bunch-Davies vacuum is expressed as
 \begin{eqnarray}
|0\rangle_{\rm BD} = N_{\gamma_p}^{-1}
\exp\left(\gamma_p\,c_R^\dagger\,c_L^\dagger\,\right)|0\rangle_{R'}|0\rangle_{L'}\,.
\label{bogoliubov3}
\end{eqnarray}
In Eq. (\ref{bogoliubov3}) we have introduced new operators $c_q=(c_R,c_L)$ that satisfy  \cite{Kanno16, Albrecht18}
\begin{eqnarray}
c_R = u\,b_R + v\,b_R^\dagger \,,\qquad
c_L = u^*\,b_L + v^*\,b_L^\dagger\,.
\label{bc}
\end{eqnarray}
The normalization factor $N_{\gamma_p}$ in Eq. (\ref{bogoliubov3}) is given by
\begin{eqnarray}
N_{\gamma_p}^2
=\left|\exp\left(\gamma_p\,c_R^\dagger\,c_L^\dagger\,\right)|0\rangle_{R'}|0\rangle_{L'}
\right|^2
=\frac{1}{1-|\gamma_p|^2}\,.
\label{norm2}
\end{eqnarray}
Considering the definition of $c_R$ and $c_L$ in Eq. (\ref{bc}) and the consistency relations from Eq. (\ref{bogoliubov3}), it is demanded that
 $c_R|0\rangle_{\rm BD}=\gamma_p\,c_L^\dag|0\rangle_{\rm BD}$,
$c_L|0\rangle_{\rm BD}=\gamma_p\,c_R^\dag|0\rangle_{\rm BD}$. Then we obtain
\begin{eqnarray}
\gamma_p = i\frac{\sqrt{2}}{\sqrt{\cosh 2\pi p + \cos 2\pi \nu}
 + \sqrt{\cosh 2\pi p + \cos 2\pi \nu +2 }}\,.
\label{gammap2}
\end{eqnarray}
For the conformally coupled
massless scalar field ($\nu=1/2$) and the minimally coupled massless scalar ($\nu=3/2$),  $\gamma_p$ simplifies to $|\gamma_p|=e^{-\pi p}$.

\section{Black-box optical parameter estimation and the GIP}\label{sec2}

\begin{figure}[htbp]
\includegraphics[height=2.8in,width=5in]{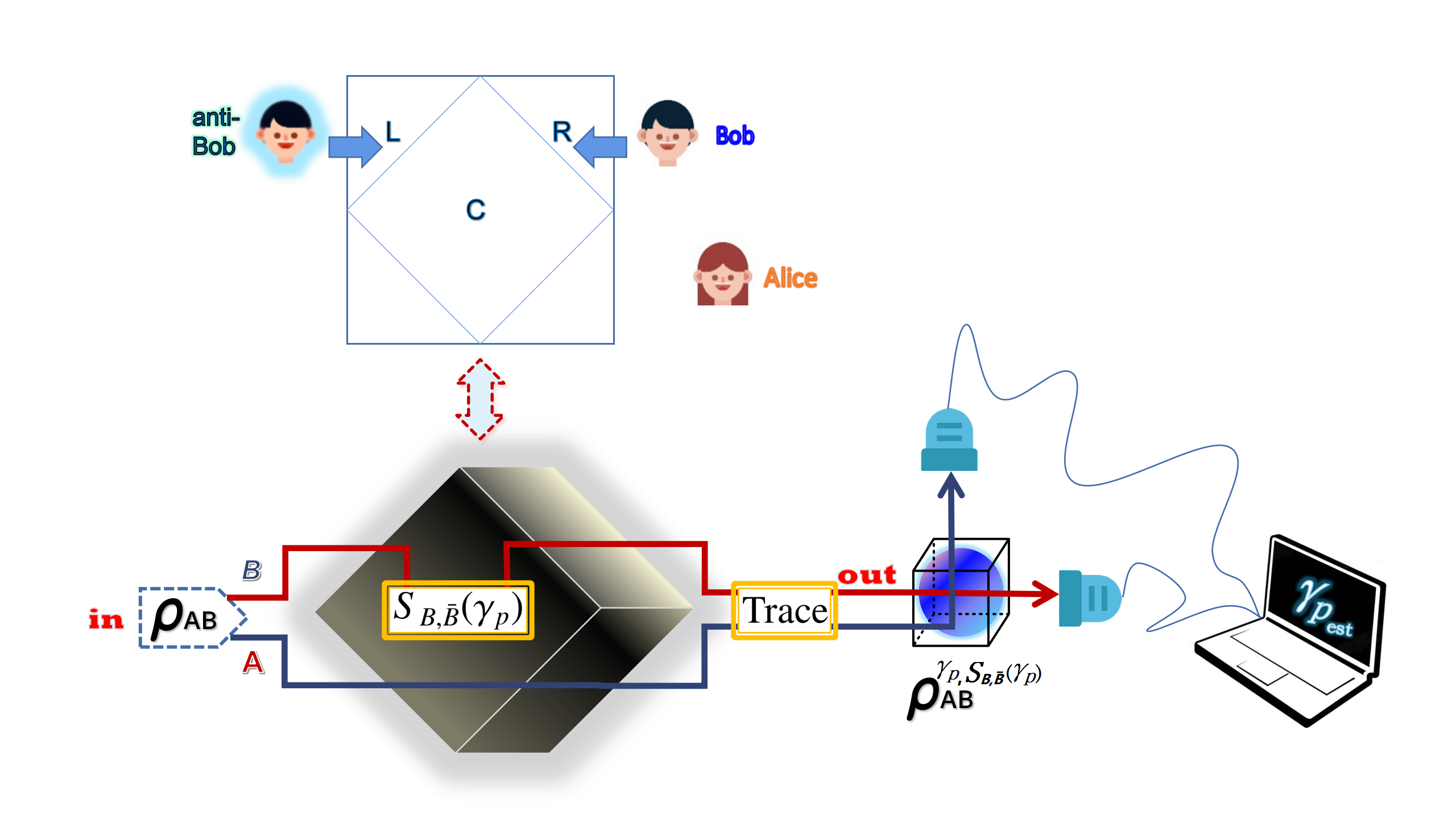}
\caption{(Color online) Schematic diagram for the  black-box optical  quantum parameter estimation in the de Sitter space.}
\label{Fig1}
\end{figure}

Our black-box parameter estimation setup is modeled as a dual-arm channel, where the estimated parameter  $\gamma_p$ is encoded to the upper arm.  As shown in Fig. 1, the proposal of black-box quantum parameter estimation includes the following steps.
Initially,  Alice and Bob share a two-mode squeezed state $\rho_{AB}$ in the
Bunch-Davies vacuum, which plays as the probe state at the ``in" side.
Then the mode $B$ in the upper arm of the channel  passes through the de Sitter space region, which acts as the black-box device.  As shown in \cite{Kanno16, Albrecht18}, under the influence of the de Sitter spacetime, the mode $B$ undergoes a two-mode squeezing transformation $ S_{B,\bar B}(\gamma_p)$, which encodes the expanding parameter $\gamma_p$ that we want to estimate.  The other mode $A$  in the lower arm of the channel  is not subjected to the spacetime. Finally, we get the output state $\rho_{AB}^{\gamma_p,S_{B,\bar B}(\gamma_p)}$ at the ``out" side by tracing over the mode $\bar B$, and one can perform a measurement on the output state $\rho_{AB}^{\gamma_p,S_{B,\bar B}(\gamma_p)}$ to construct an estimator ${\gamma_p}_{\rm est}$ for the parameter ${\gamma_p}$. If the measurements are  performed independently on the transformed state  $N$ times, the uncertainty of the parameter will be constrained by  the Cram\'er-Rao bound \cite{Braunstein1994,Cramer:Methods1946}
\begin{equation}\label{cramerrao}
N\Delta\gamma_p^2 \geq \frac{1}{{\cal F}(\rho_{AB}^{\gamma_p,S_{B,\bar B}(\gamma_p)})}\,,
\end{equation}
where the variance of the parameter $\gamma_p$ is defined as $\Delta\gamma_p^2 \equiv \langle ({\gamma_p}_{\rm est} - \gamma_p)^2\rangle$. The quantity $\cal F$ at the denominator is known as the quantum Fisher information \cite{Braunstein1994, fisher}.

On the other hand, one can define the vector of field quadratures (position and momentum) operators as $\hat{R}=(\hat{x}_{A},\hat{p}_{A},\hat{x}_{B},\hat{p}_{B})$ for a two-mode continuous-variable quantum system, which are related to the annihilation $\hat{a}_{i}$ and creation  $\hat{a}_{i}^{\dag}$
operators for each mode, by the relations $\hat{x}_{i}=\frac{(\hat{a}_{i}+\hat{a}_{i}^{\dag})}{\sqrt{2}}$
and $\hat{p}_{i}=\frac{(\hat{a}_{i}-\hat{a}_{i}^{\dag})}{\sqrt{2}i}$. The vector operator satisfies the commutation relationship: $[{{{\hat R}_i},{{\hat R}_j}} ] = i{\Omega _{ij}}$, with $\Omega  =  \bigoplus_1^{n+m} {{\ 0\ \ 1}\choose{-1\ 0}}$ being  symplectic form \cite{CM,CM2}. It is known that,
the first and second moments of a two-mode Gaussian state ${\rho _{AB}}$ can completely describe all of the properties. For the bipartite state $\rho_{AB}$, its covariance matrix (second moments) has the form \cite{CM1}
\begin{equation}\label{cm}
\sig_{AB}=\left(\begin{array}{cc}
\mathcal{A}& \mathcal{C} \\
\mathcal{C}^{\sf T} & \mathcal{B}
\end{array}\right),
\end{equation}
which can be transformed to a standard form with all diagonal $2 \times 2$ subblocks,  $\mathcal{A}=\text{diag}(a,a)$, $\mathcal{B}=\text{diag}(b,b)$, $\mathcal{C}=\text{diag}(c,d)$,
where $a,b \geq 1, c \geq |d| \geq 0$. The GIP of a two-mode Gaussian probe state with the covariance matrix $\sig_{AB}$ is defined as \cite{GIP}
\begin{equation}\label{ipcv}
{\cal P}^B(\sig_{AB}) = \frac14 \inf_{S_{B,\bar B}(\gamma_p)}{\cal F}(\sig_{AB}^{\gamma_p,S_{B,\bar B}(\gamma_p)})\,,
\end{equation}
where $\frac14$ is the normalization factor.
The GIP  ${\cal P}^B(\sig_{AB})$ represents the worst-case precision that can be obtained among all possible local dynamic choices if $\rho_{AB}$ is used as a probe state. It is a precision index that can be applied to any two-mode Gaussian state \cite{GIP}. In practice, a probe state $\rho_{AB}$ with the higher GIP reflects a more reliable metering resource, which ensures that the variance for the estimation of $\gamma_p$  smaller.

A closed formula for the GIP of two-mode Gaussian states can be obtained by introducing Eq. (\ref{cm}) into Eq. (\ref{ipcv}) \cite{GIP}
\begin{equation}\label{ipgg}
{\cal P}^B_G(\sig_{AB})=\frac{X+\sqrt{X^2+Y Z}}{2Y}\,,
\end{equation}
where
\begin{eqnarray*}
X&=&(I_{\mathcal{A}}+I_{\mathcal{C}})(1+I_{\mathcal{B}}+I_{\mathcal{C}}-I)-I^2\,, \\
Y&=&(I-1)(1+I_{\mathcal{A}}+I_{\mathcal{B}}+2I_{\mathcal{C}}+I)\,, \\
Z&=&(I_{\mathcal{A}}+I)(I_{\mathcal{A}}I_{\mathcal{B}}-I)+I_{\mathcal{C}}(2I_{\mathcal{A}}+I_{\mathcal{C}})(1+I_{\mathcal{B}})\,.
\end{eqnarray*}
In Eq. (\ref{ipgg}),  we have employed $I_{\mathcal{A}}=\det\mathcal{A}$, $I_{\mathcal{B}}=\det\mathcal{B}$, $I_{\mathcal{C}}=\det\mathcal{C}$, and $I=\det\sig_{AB}$. Here the symplectic eigenvalues
are given by $2\nu_{\pm}^2 = \Delta \pm \sqrt{\Delta^2-4I}$ with $\Delta = I_{\mathcal{A}}+I_{\mathcal{B}}+2I_{\mathcal{C}}$.

As one of the most important resources in quantum information tasks, quantum entanglement plays a significant role in quantum metrology. To better explore how to obtain higher parameter estimation accuracy, we calculate the logarithmic negativity \cite{EN1} to measure entanglement
\begin{equation}\label{lone}
\mathbf{E}_{\cal N}(\sig_{AB}) = \max\{0,\, -\ln {\tilde{\nu}_{-}}\}\,,
\end{equation}
and explore the relationship between entanglement and the GIP.
In Eq. (\ref{lone}),  the logarithmic negativity is defined in terms of  the minimum symplectic eigenvalue of the partially transposed state. Under the partial transposition, the invariant $\Delta$ is changed into $\tilde{\Delta}=I_{\mathcal{A}}+I_{\mathcal{B}}-2I_{\mathcal{C}}$  for a bipartite quantum state. The  symplectic eigenvalues are given by  $2\tilde{\nu}_{\pm}^2 = \tilde{\Delta}\pm \sqrt{\tilde{\Delta}^2-4I}$.

\section{Black-box estimation of spacetime parameters and the GIP \label{tools}}
\subsection{The worst-case precision  for Black-box metrology for the initially correlated probe state}
We assume that the mode in the black-box is observed by Bob  who resides in the open chart region R.
Initially, Bob and the global observer Alice share a two-mode squeezed state in the Bunch-Davies vacuum, which can be  described by  the covariance matrix
\begin{eqnarray}\label{inAR}
\sigma^{\rm (G)}_{AB}(s)= \left(\!\!\begin{array}{cccc}
\cosh (2s)I_2&\sinh (2s)Z_2\\

\sinh (2s)Z_2&\cosh (2s)I_2\\
\end{array}\!\!\right),
\end{eqnarray}
where $s$ is the squeezing of the  initial state and $I_2= {{\ 1\ \ 0}\choose{\ 0\ \ 1}},$ $Z_2= {{\ 1\ \ 0}\choose{\ 0 \  -1}}$.
As showed in \cite{Kanno16}, the Bunch-Davies vacuum for a global observer  can be expressed as the squeezed state of the $R$ and $L$ vacua
\begin{eqnarray}
\nonumber|0\rangle_{\rm BD}=\sqrt{1-|\gamma_p|^2}\,\sum_{n=0}^\infty\gamma_p
^n|n\rangle_L|n\rangle_R\,,
\label{bogoliubov2}
\end{eqnarray}
where $\gamma_B$ is the  squeezing parameter given in Eq. (\ref{gammap2}). In the phase space, we use a symplectic operator $ S_{B,\bar B}(\gamma_p)$ to express  such transformation, which is
\begin{eqnarray}\label{cmtwomode}
 S_{B,\bar B}(\gamma_p)= \frac{1}{\sqrt{1-|\gamma_p|^2}}\left(\!\!\begin{array}{cccc}
1&0&|\gamma_p|&0\\
0&1&0&-|\gamma_p|\\
|\gamma_p|&0&1&0\\
0&-|\gamma_p|&0&1
\end{array}\!\!\right),
\end{eqnarray}
where $ S_{B,\bar B}(\gamma_p)$  denotes that the squeezing transformation is performed to the bipartite state shared between Bob and anti-Bob ($\bar B$).

Bob's observed mode is mapped into two open charts under this transformation. That is to say, an extra set of modes $\bar{B}$ is relevant from the perspective of a  observer in the open charts. Then we can calculate the covariance matrix of the entire state\cite{adesso3}, which is
\begin{eqnarray}\label{All3}
\sigma^{\rm }_{AB \bar B}(s,\gamma_p)  &=& \left(
       \begin{array}{ccc}
          \mathcal{\sigma}_{A} & \mathcal{E}_{AB} & \mathcal{E}_{A\bar B} \\
         \mathcal{E}^{\sf T}_{AB} &  \mathcal{\sigma}_{B} & \mathcal{E}_{B\bar B} \\
         \mathcal{E}^{\sf T}_{A\bar B} & \mathcal{E}^{\sf T}_{B\bar B} &  \mathcal{\sigma}_{\bar B} \\
       \end{array}
     \right)
 \,,
\end{eqnarray}
where $\sigma^{\rm (G)}_{AB}(s) \oplus I_{\bar B}$ is the initial covariance matrix for the entire system. In the above expression, the diagonal elements are given by
\begin{equation} \mathcal{\sigma}_{A}=\cosh(2s)I_2,\end{equation}
 \begin{equation}
\mathcal{\sigma}_{B}=\frac{|\gamma_p|^2+\cosh(2s)}{1-|\gamma_p|^2}I_2,
\end{equation} and
\begin{equation}
\mathcal{\sigma}_{\bar B}=\frac{1+|\gamma_p|^2\cosh(2s)}{1-|\gamma_p|^2}I_2.
 \end{equation}
Similarly, we find that the non-diagonal elements are
$\mathcal{E}_{AB}=\frac{\sinh(2s)}{\sqrt{1-|\gamma_p|^2}}Z_2$, $\mathcal{E}_{B\bar B}=\frac{|\gamma_p|(\cosh(2s)+1)}{1-|\gamma_p|^2}Z_2$ and $\mathcal{E}_{A\bar B}=\frac{|\gamma_p|\sinh(2s)}{\sqrt{1-|\gamma_p|^2}}I_2$.

Because Bob in  chart $R$ has no access to the modes in the causally disconnected region $L$, we must trace over the inaccessible modes. Then one obtains
covariance matrix $\sigma_{AB}$ for Alice and Bob
\begin{eqnarray}\label{AB}
\sig_{AB}(s,\gamma_p)= \left(\!\!\begin{array}{cccc}
\cosh(2s)&0&\frac{\sinh(2s)}{\sqrt{1-|\gamma_p|^2}}&0\\
0&\cosh(2s)&0&-\frac{\sinh(2s)}{\sqrt{1-|\gamma_p|^2}}\\
\frac{\sinh(2s)}{\sqrt{1-|\gamma_p|^2}}&0&\frac{|\gamma_p|^2+\cosh(2s)}{1-|\gamma_p|^2}&0\\
0&-\frac{\sinh(2s)}{\sqrt{1-|\gamma_p|^2}}&0&-\frac{|\gamma_p|^2+\cosh(2s)}{-1+|\gamma_p|^2}
\end{array}\!\!\right).
\end{eqnarray}
From Eq. (\ref{AB}), we obtain $\Delta^{(AB)}=1+\frac{(1+|\gamma_p|^2\cosh(2s))^2}{(1-|\gamma_p|^2)^2}$.
The GIP and entanglement of this state can be calculated via Eq. (\ref{ipgg}) and Eq. (\ref{lone}).

\begin{figure}[htbp]
\centering
\includegraphics[height=2.2in,width=2.8in]{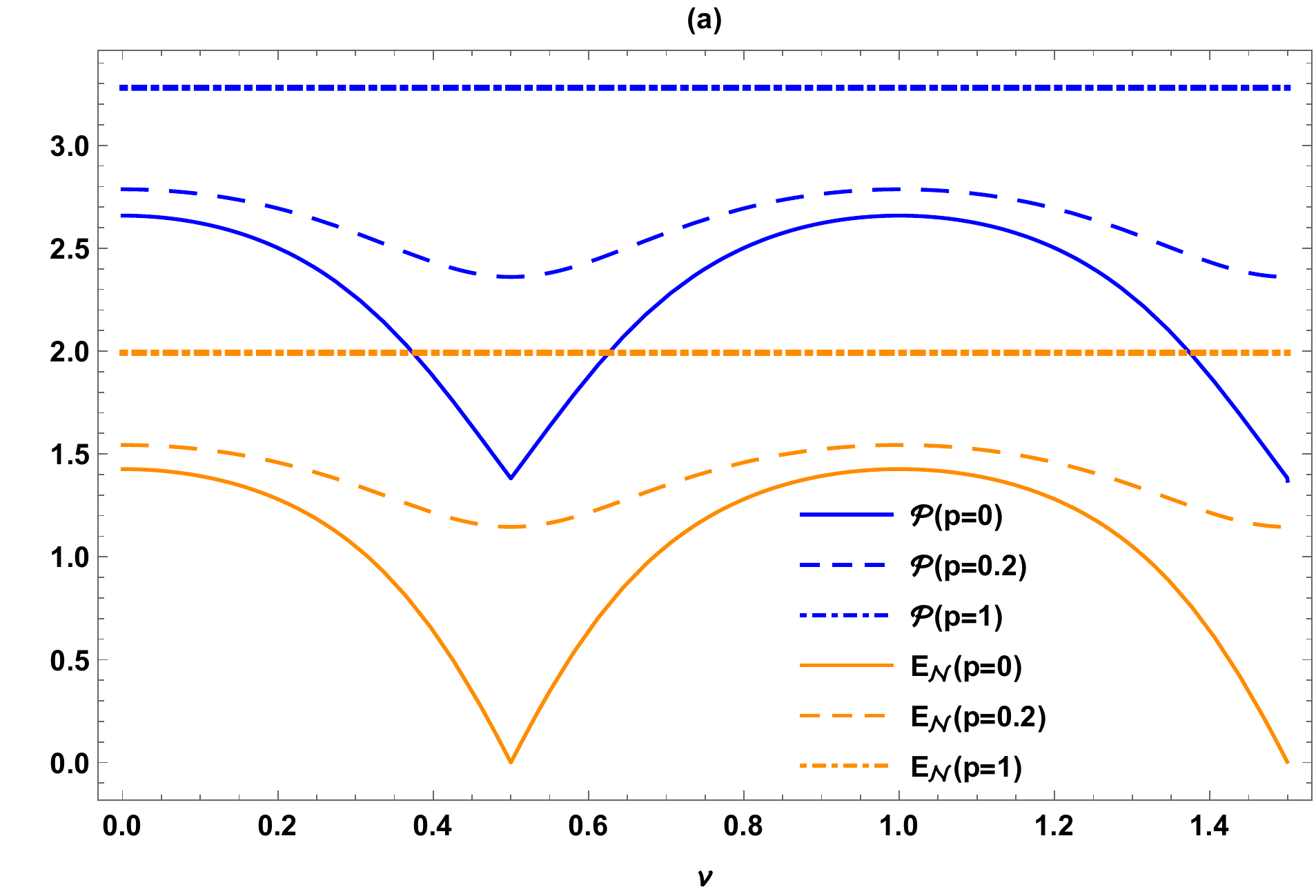}
\includegraphics[height=2.2in,width=2.8in]{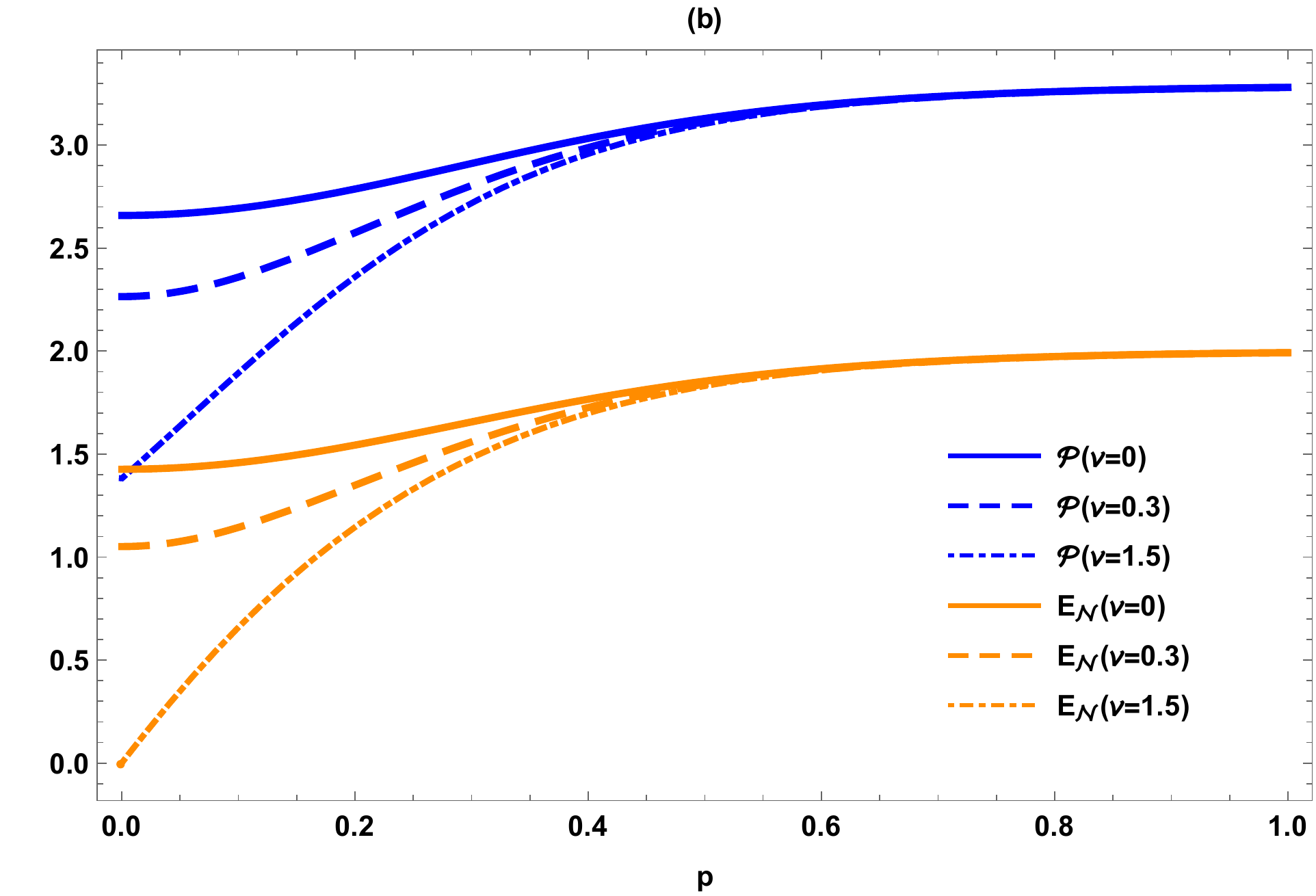}
\caption{ (Color online) (a) The GIP and quantum entanglement between Alice and Bob versus  the mass parameter $\nu$. (b) The evolution of the GIP and quantum entanglement versus the space curvature parameter $p$. The initial squeezing parameter is fixed as $s=1$.  }\label{Fig1}
\end{figure}

In Fig. 2(a) and Fig. 2(b), we demonstrate the function images of the GIP and quantum entanglement for the state $\sig_{AB}(s,\gamma_p)$ as a function of the mass parameter $\nu$ and the space curvature parameter $p$, respectively. It can be seen in Fig. 2 (a) that the variation trend of the GIP and entanglement is basically the same, which verifies the fact that quantum entanglement is the resource of parameter estimation. As mentioned above, the GIP quantifies the precision for the estimation of encoded parameters in a worst-case scenario.  A probe state with the higher GIP reflects  more reliable metering resources, which ensures the variance for the estimated parameter  is smaller. Therefore,  we can conclude that the precision of the black-box quantum metrology depends on the abundance of the quantum entanglement  in the de Sitter space.
It is interesting to find that, either the precision of the black-box parameter estimation or the resource of entanglement is unaffected by the mass parameter $\nu$ in the flat space limit ($p=1$). However, the mass parameter of the field has remarkable effects on the GIP as well as the  accuracy of black-box metrology in the curved de Sitter space.  In particular, both of them reach non-zero minimum values in the conformal scalar limit  ($\nu=1/2$) and massless scalar limit ($\nu=3/2$).

In Fig. 2 (b), we can see that both the GIP and entanglement are monotone increasing functions of the curvature parameter $p$. Considering that  the effect of the curvature becomes stronger when $p$ becomes less and less from $1$ to $0$  \cite{Kanno16, Sasaki:1994yt,Albrecht18,DE1,DE2,DE3},  this in fact demonstrates that the lower the space curvature, the higher the accuracy of parameter estimation.  It is worth noting that the values of the GIP and entanglement are more sensitive to the spacetime curvature  in the massless scalar limit  $\nu=3/2$. Conversely, more quantum entanglement is reserved and higher precision can be attained for the case $\nu=0$.

\subsection{The dynamics of GIP between initially unrelated modes}

We also interested in whether the bipartite state between the initially uncorrelated modes can be employed as probe state for the black-box estimation. To this end we study the behavior of the quantum parameter estimation between all the bipartite pairs in the de Sitter spacetime.
The covariance matrix $\sigma_{A\bar B}$ between the observer Alice and the other observer anti-Bob in the region $L$ is obtained by tracing over the modes $B$
\begin{eqnarray}\label{CM2}
\sigma_{A\bar B}(s,\gamma_p)= \left(\!\!\begin{array}{cccc}
\cosh(2s)&0&\frac{|\gamma_p| \sinh(2s)}{\sqrt{1-|\gamma_p|^2}}&0\\
0&\cosh(2s)&0&\frac{|\gamma_p| \sinh(2s)}{\sqrt{1-|\gamma_p|^2}}\\
\frac{|\gamma_p|\sinh(2s)}{\sqrt{1-|\gamma_p|^2}}&0&\frac{1+|\gamma_p|^2\cosh(2s)}{1-|\gamma_p|^2}&0\\
0&\frac{|\gamma_p|\sinh(2s)}{\sqrt{1-|\gamma_p|^2}}&0&\frac{1+|\gamma_p|^2\cosh(2s)}{1-|\gamma_p|^2}
\end{array}\!\!\right).
\end{eqnarray}
The GIP and entanglement of state $\sigma_{A\bar B}$ can be obtained by using this covariance matrix and have been plotted in Fig. 3.

\begin{figure}[htbp]
\includegraphics[height=2.2in, width=2.8in]{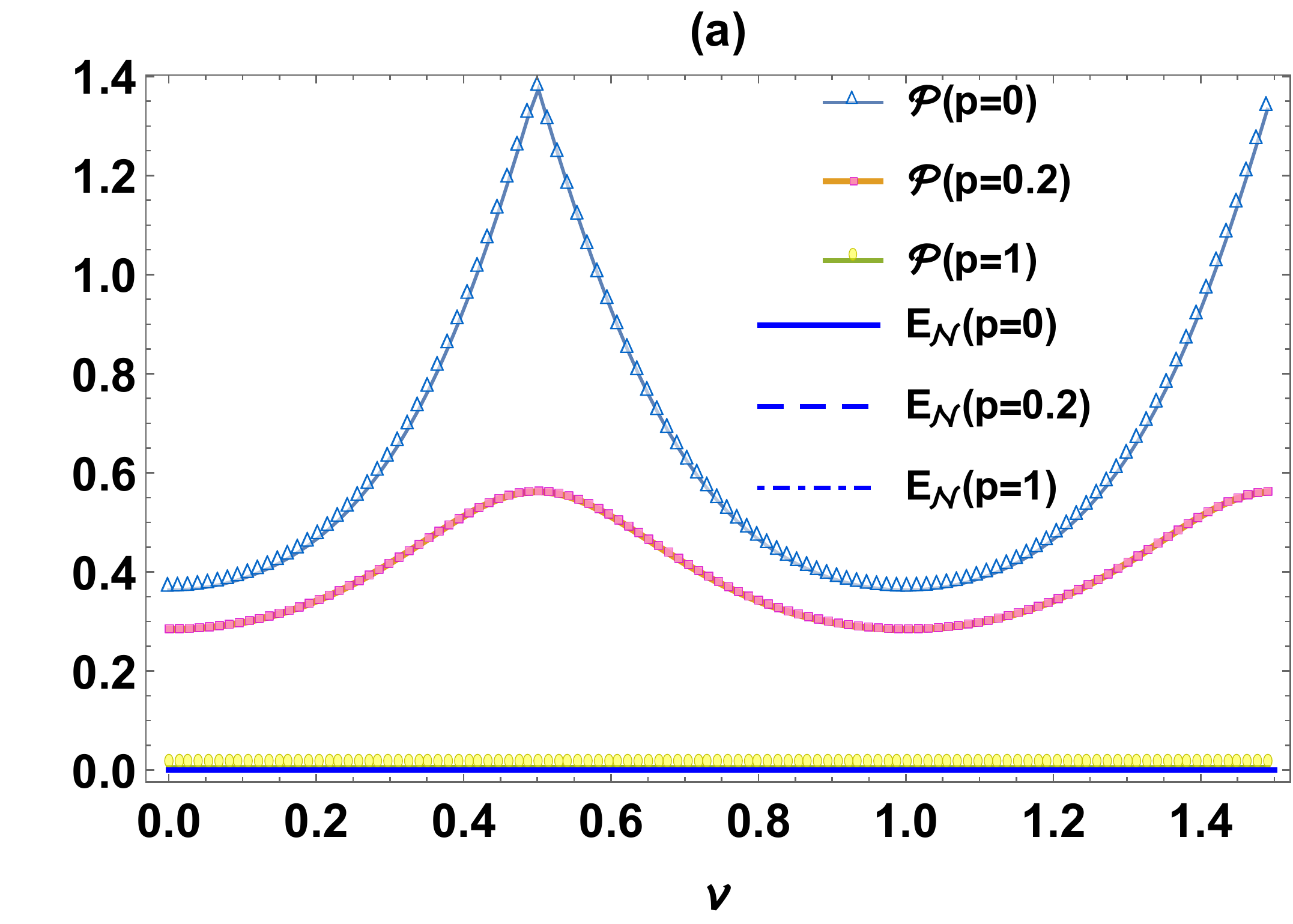}
\includegraphics[height=2.2in, width=2.8in]{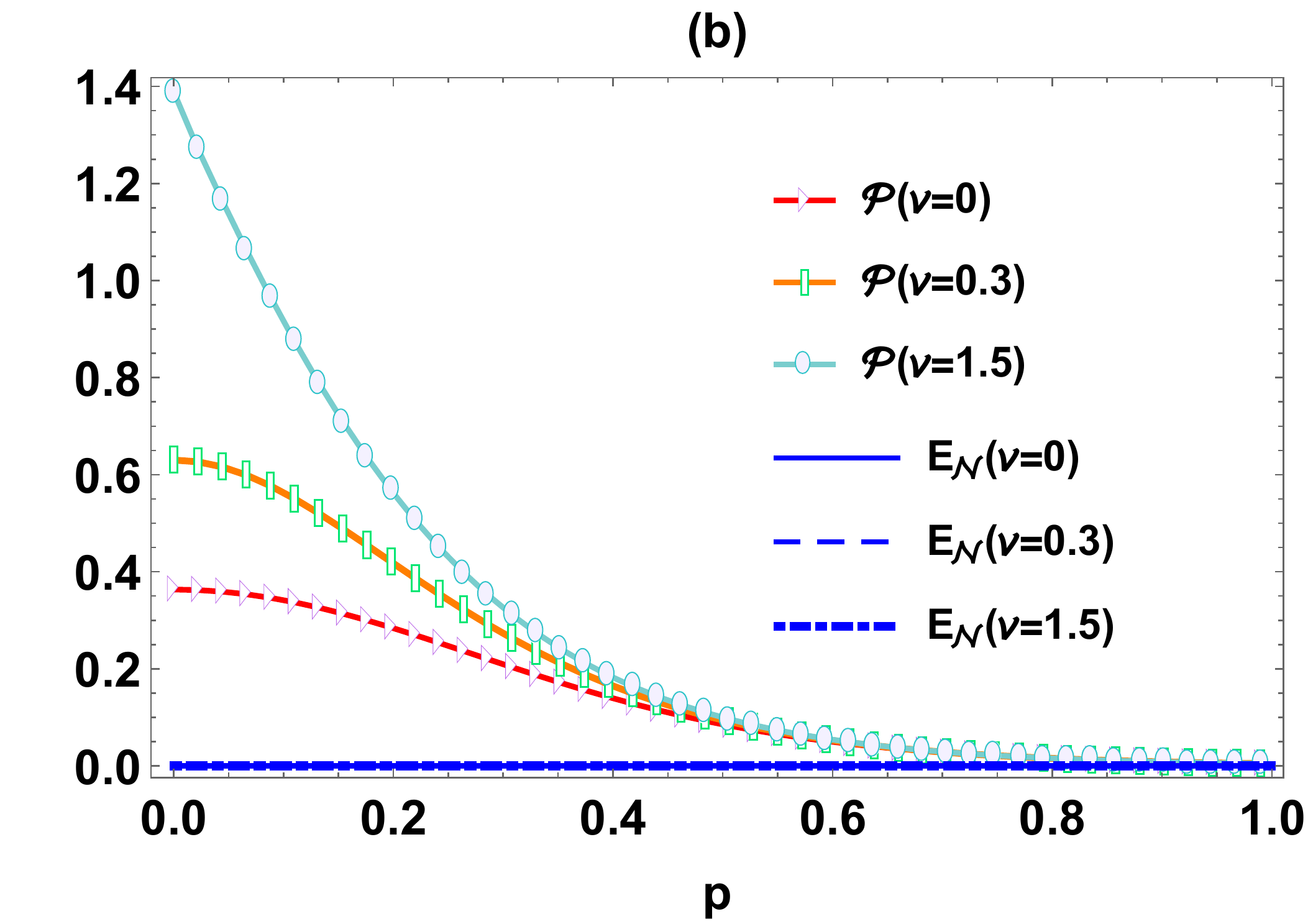}
\caption{ (Color online) (a) Plots of the GIP ($\mathcal{P}$) and the quantum entanglement ($\mathbf{E}_{\cal N}$) for  the probe state  $\rho_{A\bar B}$ as a function of the mass parameter $\nu$.  (b) Quantum entanglement and the GIP vs. the variation of the space curvature parameter $p$. The initial squeezing parameter is fixed as $s=1$.  }\label{Fig2}
\end{figure}

In Fig. 3, we can see that the quantum entanglement between Alice and anti-Bob is zero for any different mass and curvature parameters. That is to say, the curvature of the de Sitter spacetime dose not generate quantum entanglement between Alice and anti-Bob \cite{DE1}.  It is shown from Fig. 3(a) that the mass of the scalar field remarkably affects the accuracy of  the black-box parameter estimation in the de Sitter space, which is quite different from the flat space case ($p=1$) where the mass parameter  does not influence the precision of  estimation.   If the space curvature parameter is less than $1$, the GIP is generated  by the curvature effect of the de Sitter space. This means the advantage of quantum metrology is ensured in the curved spacetime even the entanglement is zero.
In addition, the GIP reaches its maximum when the mass parameters are $\nu=1/2$ (conformal scalar limit) and $\nu=3/2$ (massless scalar limit).
In Fig. 3(b), it is shown that for the probe state $\rho_{A\bar B}$, the GIP increases with the increase of curvature effect. Again, we find that the GIP always maintains a non-zero value if the space curvature parameter is less than $1$. This means the precision of the black-box estimation is enhanced under the influence of the de Sitter curvature. In other words, the role of quantum resources for the black-box estimation is played by a non-entanglement quantum correlation. The spacetime effect of the expanding universe may generate some non-entanglement   quantum resources between Alice and Bob.

The discord-type  quantum correlation \cite{zurek} is believed to be more practical than entanglement to
describe the quantum resources in  certain quantum systems. As seen in Refs.
\cite{experimental-onequbit}, although there is no
entanglement,  quantum information processing tasks can also
be done efficiently. We wonder if  quantum discord is generated by  spacetime effects of the de Sitter space.
To this end, we introduce the Gaussian quantum discord  \cite{discordGIP} to explore the relationship between the GIP and discord-type quantum correlation . For the two-mode Gaussian state $\rho_{AB}$, the R\'{e}nyi-$2$ measure of quantum discord ${\cal D}_2(\rho_{A|B})$ admits the following expression \cite{discord,CM2}
\begin{eqnarray}\label{eq}
{\cal D}_2(\rho_{A|B}) &=& \ln b - \frac12 \ln \big(\det{\sig_{AB}}\big) +   \frac12 \ln \left(\inf_{\lambda,\varphi}{\det\tilde{\gr\sigma}^{\Pi_{\lambda,\varphi}}_A}\right)\,. \label{eq:D2AB}
\end{eqnarray}
In particular, for the form covariance matrix given in Eq. (\ref{cm}), one obtains
\begin{eqnarray} \label{eq:optdetcond}
&\inf_{\lambda,\varphi}\det\tilde{\gr\sigma}^{\Pi_{\lambda,\varphi}}_A = \left\{
\begin{array}{l}
 a \left(a-\frac{c_+^2}{b}\right)\,, \\
\text{if}{\scriptstyle{\left(a b^2 c_-^2-c_+^2 \left(a+b c_-^2\right)\right) \left(a b^2 c_+^2-c_-^2 \left(a+b c_+^2\right)\right)<0}}\, \mbox{;} \\ \quad \\
 \scriptscriptstyle{\frac{2 \left|c_- c_+\right| \sqrt{\left(a \left(b^2-1\right)-b c_-^2\right) \left(a \left(b^2-1\right)-b c_+^2\right)}+\left(a \left(b^2-1\right)-b c_-^2\right) \left(a \left(b^2-1\right)-b c_+^2\right)+c_-^2 c_+^2}{\left(b^2-1\right)^2}} \,, \\
\text{otherwise.}
\end{array}
\right.
\end{eqnarray}
Inserting Eq.~(\ref{eq:optdetcond}) into Eq.~(\ref{eq}), we can compute  the  discord of the  two-mode Gaussian state. 

\begin{figure}[htbp]
\includegraphics[height=2.6in,width=3.4in]{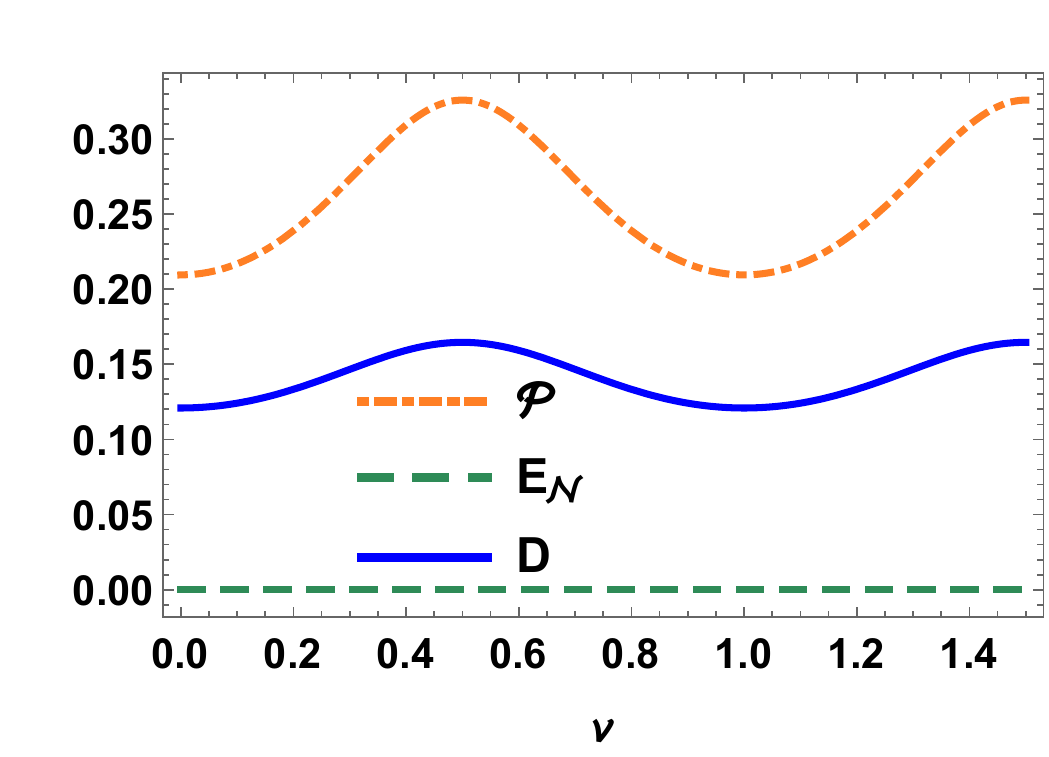}
\caption{ (Color online).  The variation functions of quantum discord (blue solid line), the GIP (orange dotted line) and entanglement (green dotted line) between Alice and anti-Bob vs. the mass parameter of the field. Other parameters are fixed as  $s = 1$ and  $p=0.3$. }\label{Fig4}
\end{figure}

 In Fig. 4, we plot the  quantum discord, the GIP as well as the entanglement between  Alice and anti-Bob as a function of the mass parameter.  It is shown that the quantum-enhance parameter is ensured for the probe state $\rho_{A\bar B}$ because the GIP is always nonzero. The variation trend of the GIP  is in accord with the quantum discord. This means  the quantum discord in the probe state provides quantum resources for the black-box quantum parameter estimation. It is interesting to note that the bipartite state between Alice and anti-Bob can also be employed as the probe state for the black-box estimation in the expanding de Sitter spacetime even if there is no entanglement generated between them. Therefore, discord-type quantum correlation for the input probe is the key resource for black-box estimation in the de Sitter space, which ensures the advantage of quantum metrology.

We are also interested in the behavior of the GIP between Bob and anti-Bob, which are separated by the event horizon of the de Sitter space. We get the covariance matrix $\sigma_{B\bar B}$ by tracing off the mode $A$
\begin{eqnarray}\label{AbarB}
\sigma_{B\bar B}(s,\gamma_p)= \left(\!\!\begin{array}{cccc}
\frac{|\gamma_p|^2+\cosh(2s)}{1-|\gamma_p|^2}&0&\frac{2|\gamma_p|\cosh^2(s)}{1-|\gamma_p|^2}&0\\
0&\frac{|\gamma_p|^2+\cosh(2s)}{1-|\gamma_p|^2}&0&-\frac{2|\gamma_p|\cosh^2(s)}{1-|\gamma_p|^2}\\
\frac{2|\gamma_p|\cosh^2(s)}{1-|\gamma_p|^2} &0&\frac{1+|\gamma_p|^2\cosh(2s)}{1-|\gamma_p|^2}&0\\
0&-\frac{2|\gamma_p|\cosh^2(s)}{1-|\gamma_p|^2} &0&\frac{1+|\gamma_p|^2\cosh(2s)}{1-|\gamma_p|^2}
\end{array}\!\!\right).
\end{eqnarray}
Similarly, the entanglement and the GIP of Bob and anti-Bob can be calculated by employing the measurement introduced above.

\begin{figure}[htbp]
\includegraphics[height=2.1in,width=2.8in]{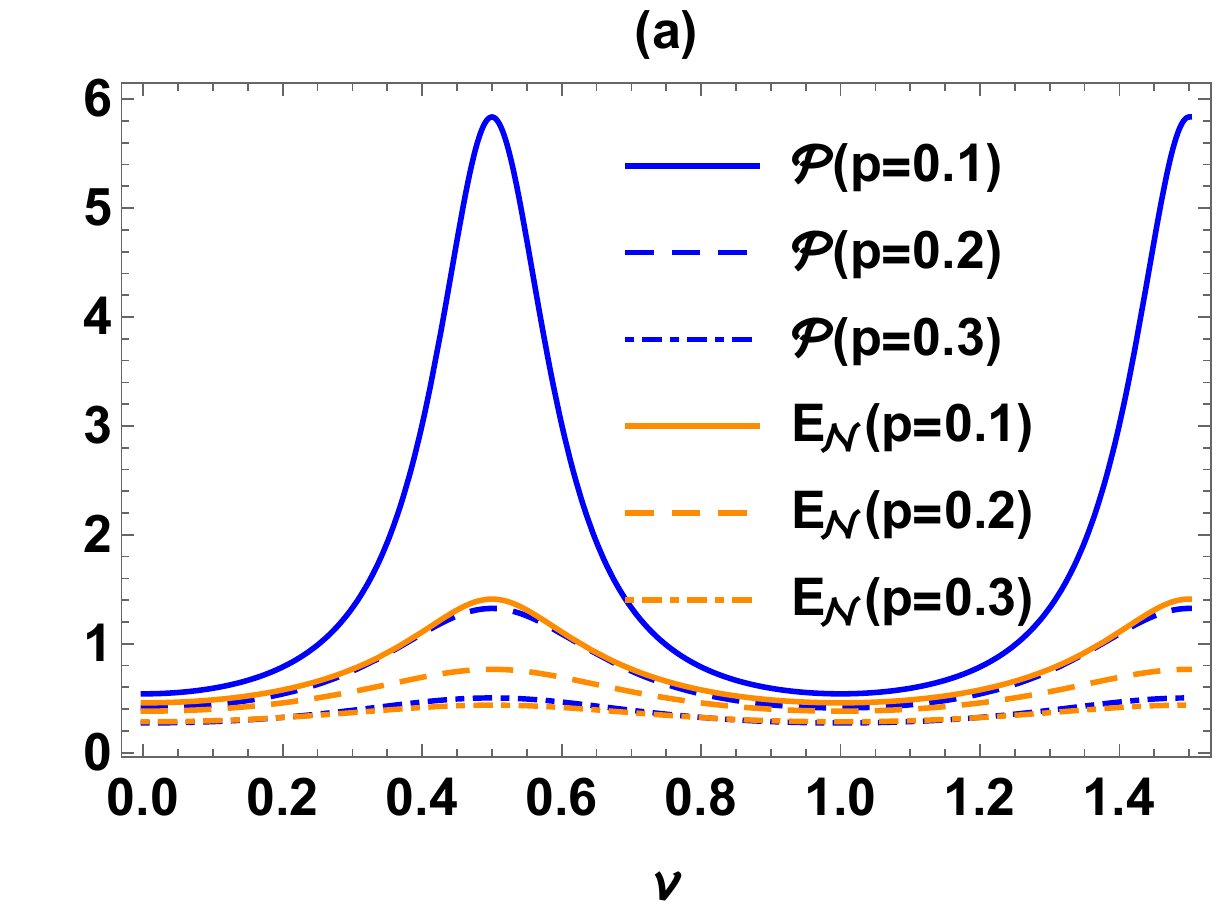}
\includegraphics[height=2.1in,width=2.8in]{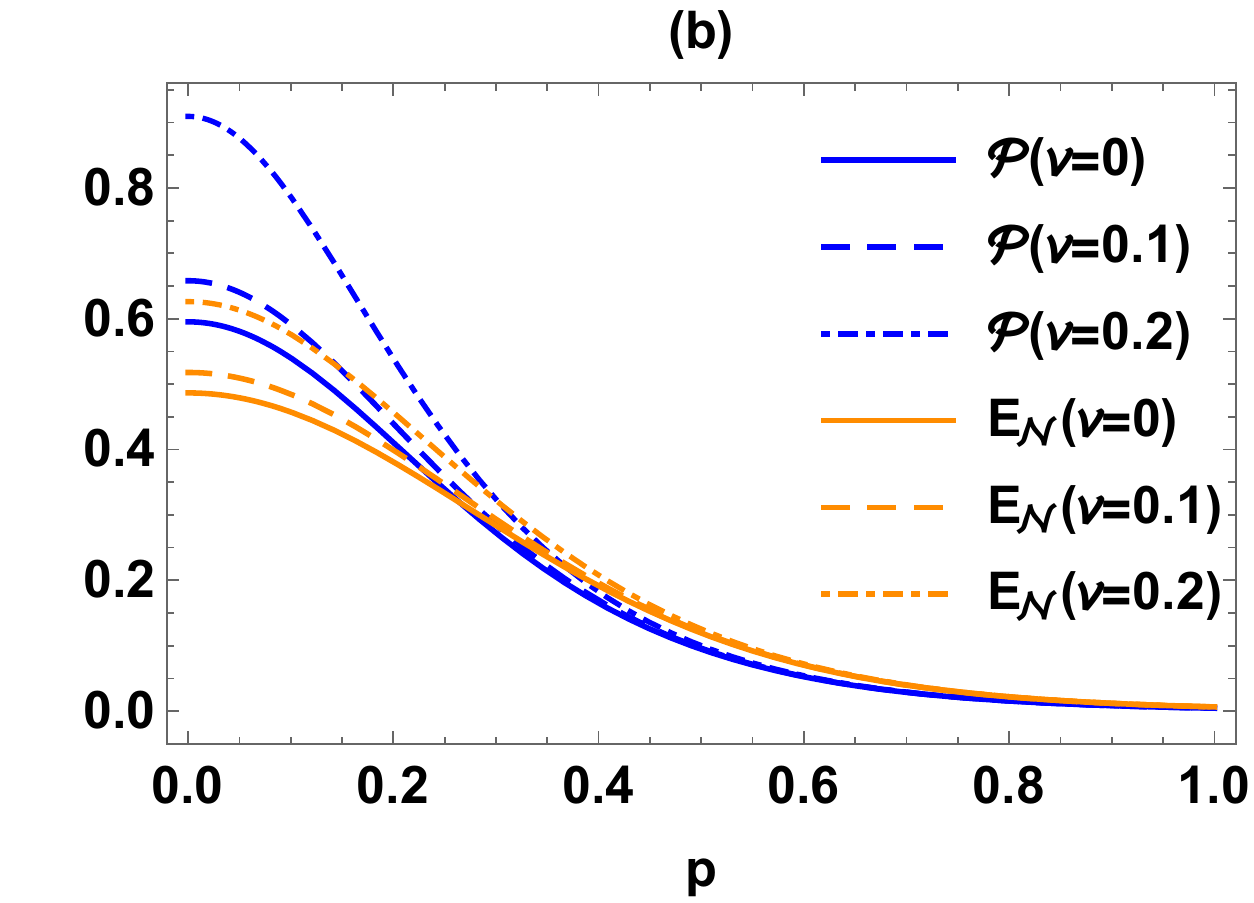}
\caption{ (Color online) (a) Plots of the GIP ($\mathcal{P}$) and the quantum entanglement ($\mathbf{E}_{\cal N}$)  for  the probe state  $\rho_{B\bar B}$  as a function of the mass parameter $\nu$ of the field. (b) Plots of the GIP ($\mathcal{P}$) and the quantum entanglement ($\mathbf{E}_{\cal N}$) as a function of the space curvature parameter $p$. The initial squeezing parameter  is fixed as  $s=1$.}\label{Fig5}
\end{figure}

In Fig. 5, we plot the behavior of the GIP and the entanglement between Bob and anti-Bob. It is shown that the GIP  increases significantly when the curvature  parameter  $p\to0$, i.e.,  the effect of curvature gets stronger. In fact, this indicates  space curvature in the de Sitter space enhances the precision of quantum parameter estimation if $\rho_{B\bar B}$ is employed as the probe state.   This is nontrivial because the open charts $R$ and $L$ are causally disconnected in the de Sitter spacetime
and classical information can not be exchanged between the causally disconnected regions according to the general relativity.
Here we find that quantum precision measurements are achieved between causally disconnected areas in the de Sitter space, and a more accurate experimental result can be performed in a more  curved spacetime. Unlike the initial correlation system Alice and Bob, the GIP and entanglement are maximized for the mass parameters $\nu=1/2$ (conformal) and $\nu=3/2$ (massless). This is because the curvature effect of the de Sitter space damages the entanglement and the GIP for the probe state $\rho_{AB}$, while it generates entanglement for the probe state $\rho_{B\bar B}$.

\section{Conclusions}
We have proposed a  black-box quantum parameter estimation scheme for the expanding parameter and discussed the behavior of the GIP for the de Sitter space.
It is found that  under the curvature effect of the de Sitter space, the changes of the GIP for the probe states $\rho_{AB}$ and $\rho_{B \bar B}$  are consistent with the changes of entanglement.
This verifies the fact that quantum resources provide a guarantee for obtaining the higher GIP, and the probe state with the higher GIP ensures a  smaller variance for the estimation of expanding parameter.
It is demonstrated that the change of the mass parameter does not affect the minimum accuracy of black-box parameter estimation in flat space,
but has remarkable effects on the GIP of the black-box parameter estimation in the curved space. Interestingly,  the state between separated open charts can be employed as the probe state for the black-box quantum metrology.
It is worth noting that the quantum discord of probe states serves as a promising resource for the black-box quantum parameter estimation when there is no entanglement between the initially uncorrelated open charts.
The behavior of the probe state for the black-box estimation is quite different because the curvature effect of the de Sitter space damages the entanglement and the GIP for the initially correlated probe state, while it generates quantum resources  for initially uncorrelated probe states.

\begin{acknowledgments}
This work is supported by the National Natural Science Foundation of China under Grant No. 12122504 and  No.11875025.
\end{acknowledgments}


\begin{thebibliography}{99}
\bibitem{advances}
 V. Giovannetti, S. Lloyd, and L. Maccone, Nat. Photon.  \textbf{5}, 222 (2011).

\bibitem{EN1}
W. K. Wootters, Phys. Rev. Lett. {\bf 80}, 2245 (1998).


\bibitem{discord1}
 H. Ollivier and W. H. Zurek, Phys. Rev. Lett. {\bf 88}, 017901 (2001).


\bibitem{Braunstein1994}
S. L. Braunstein and C. M. Caves, Phys. Rev. Lett. \textbf{72}, 3439 (1994).

\bibitem{Grote}
H. Grote {\it et al.},  Phys. Rev. Lett.  \textbf{110}, 181101(2013).



\bibitem{Aasi2013}
J. Aasi {\it et al.}, Nat. Photon.  \textbf{7}, 613 (2013).

\bibitem{Ma2017}
Y. Ma {\it et al.},  Nat. Phys.  \textbf{13}, 776 (2017).


\bibitem{Guo2020}
X.~Guo {\it et al.}, Nat. Phys. \textbf{16}, 281 (2020).

\bibitem{zhao2021}
S. R. Zhao {\it et al.}, Phys. Rev. X \textbf{11}, 031009 (2021).

\bibitem{kolobov1999spatial}
M.~I. Kolobov, Rev. Mod. Phys. \textbf{71}, 1539 (1999).

\bibitem{qi1}
C. Lupo, Z. Huang, and P. Kok, Phys. Rev. Lett. {\bf 124}, 080503 (2020).


\bibitem{leibfried2004toward}
D.~Leibfried {\it et al.}, Science \textbf{304}, 1476 (2004).

\bibitem{wangsyn}
J. Wang, Z.  Tian, J. Jing, and H. Fan, Phys. Rev. D \textbf{93}, 065008 (2016).

\bibitem{Rquan}
R. Quan {\it et al.}, Optics Letters \textbf{44}, 000614 (2019).

\bibitem{aspachs}
M. Aspachs, G. Adesso, and I. Fuentes, Phys. Rev. Lett. \textbf{ 105}, 151301 (2010).



\bibitem{HoslerKok2013}
D. Hosler and P. Kok, Phys. Rev. A \textbf{88}, 052112 (2013).

 \bibitem{RQI6}
Y. Huang, K. Yan, Y. Wu, X. Hao, Eur. Phys. J. C {\bf 79}, 11 (2019).

\bibitem{RQM}
M. Ahmadi, D. E. Bruschi, C. Sab\'in, G. Adesso, and I. Fuentes, Sci. Rep. \textbf{4}, 4996 (2014).

\bibitem{RQM2}
M. Ahmadi, D. E. Bruschi, and I. Fuentes, Phys. Rev. D \textbf{ 89}, 065028 (2014).

\bibitem{detector1}
J. Wang, L. Zhang, S. Chen, J. Jing, Phys. Lett. B {\bf 802}, 135239 (2020).

\bibitem{detector2}
X.  Liu, J. Jing, Z. Tian, W. Yao, Phys. Rev. D {\bf 103}, 125025 (2021).

\bibitem{jieciRQM1}
J. Wang, Z. Tian, J. Jing, and H. Fan, Nucl. Phys. B \textbf{892}, 390 (2015).

\bibitem{RQMuiverse2}
H. Du and  R. B. Mann, JHEP {\bf 05}, 112 (2021).


\bibitem{RQM8}
D. E. Bruschi, A. Datta, R. Ursin, T. C. Ralph, and I. Fuentes,  Phys. Rev. D \textbf{90}, 124001 (2014).

\bibitem{RQM9}
S. P. Kish and T. C. Ralph, Phys. Rev. D \textbf{99}, 124015 (2019).



\bibitem{JHEP1}
J. Doukas, S. Y. Lin, B.L. Hu, and R.B. Mann, JHEP {\bf 11}, 119 (2013).

\bibitem{JHEP2}
P. H. Liu and F. L. Lin, JHEP {\bf 07}, 084 (2016).

\bibitem{JHEP3}
S. Banerjee, A. K. Alok, S. Omkar, and R. Srikanth, JHEP {\bf 02}, 082 (2017).

\bibitem{discordGIP}
 D. Girolami {\it et al.}, Phys. Rev. Lett. {\bf 112}, 210401 (2014).


\bibitem{GIP}
G. Adesso, Phys. Rev. A {\bf 90}, 022321 (2014).

\bibitem{Sasaki:1994yt}
M.~Sasaki, T.~Tanaka, and K.~Yamamoto,
  Phys.\ Rev.\ D {\bf 51}, 2979 (1995).

\bibitem{Kanno16}
S. Kanno,  J.~P.~Shock, and J.~Soda,
  Phys.\ Rev.\ D {\bf 94}, 125014 (2016).


\bibitem{Albrecht18}
A. Albrecht, S. Kanno, and M. Sasaki,
  Phys.\ Rev.\ D {\bf 97}, 083520 (2018).



\bibitem{DE1}
J. Wang, C. Wen, J. Jing, and S. Chen, Phys. Lett. B  {\bf800}, 135109 (2020).

\bibitem{DE2}
C. Wen, J. Wang, and J. Jing, Eur. Phys. J. C  {\bf80}, 78 (2020).

\bibitem{DE3}
Q. Liu, C. Wen, J. Wang, and J. Jing, Ann. Phys. {\bf533}, 2000536 (2021).


\bibitem{bov3}
J. Maldacena and G.L. Pimentel, JHEP {\bf1302}, 038 (2013).
\bibitem{Cramer:Methods1946}
H.~Cram$\mathrm{\acute{e}}$r,\
\textit{Mathematical Methods of Statistics}\
(Princeton University, Princeton, NJ, 1946).


\bibitem{fisher}
M. G. A. Paris, Int. J. Quant. Inf. \textbf{07}, 125 (2009).

\bibitem{CM}
G. Adesso and F. Illuminati, J. Phys. A \textbf{40}, 7821 (2007).


\bibitem{CM2}
G. Adesso, S. Ragy, and A. R. Lee, Open Syst. Inf. Dyn. \textbf{21}, 1440001 (2014).



\bibitem{CM1}
C. Weedbrook {\it et al.}, Rev. Mod. Phys. \textbf{84}, 621 (2012).


 \bibitem{adesso3}
G. Adesso, I. Fuentes-Schuller, and M. Ericsson, Phys. Rev. A {\bf 76}, 062112 (2007).



\bibitem {zurek}
 H. Ollivier and W. H. Zurek, Phys. Rev. Lett. {\bf 88}, 017901 (2001).

\bibitem {experimental-onequbit}
B. P. Lanyon, M. Barbieri, M. P. Almeida, and A. G. White, Phys. Rev. Lett.{\bf 101}, 200501 (2008).

 \bibitem{discord}
G. Adesso and A. Datta, Phys. Rev. Lett. {\bf 105}, 030501 (2010).























\end{thebibliography}
\end{document}